\documentclass[%
reprint,
 amsmath,amssymb,
aps,
groupedaddress,
]{revtex4-1}

\usepackage{graphicx}
\usepackage{dcolumn}
\usepackage{bm}

\begin{document}


\title{Pure spin current transport in a SiGe alloy}

\author{T. Naito,$^{1}$ M. Yamada,$^{1}$ M. Tsukahara,$^{1}$ S. Yamada,$^{1,2}$ K. Sawano,$^{3}$ and K. Hamaya$^{1,2}$\footnote{E-mail: hamaya@ee.es.osaka-u.ac.jp}}

\affiliation{
$^{1}$Department of Systems Innovation, Graduate School of Engineering Science, Osaka University, 1-3 Machikaneyama, Toyonaka 560-8531, Japan.}
\affiliation{
$^{2}$Center for Spintronics Research Network, Graduate School of Engineering Science, Osaka University, 1-3 Machikaneyama, Toyonaka 560-8531, Japan.}
\affiliation{
$^{3}$Advanced Research Laboratories, Tokyo City University, 8-15-1 Todoroki, Tokyo 158-0082, Japan.
}

\date{\today}

\begin{abstract}
Using four-terminal nonlocal magnetoresistance measurements in lateral spin-valve devices with Si$_{\rm 0.1}$Ge$_{\rm 0.9}$, we study pure spin current transport in a degenerate SiGe alloy ($n \sim$ 5.0 $\times$ 10$^{18}$ cm$^{-3}$). 
Clear nonlocal spin-valve signals and Hanle-effect curves, indicating generation, transport, and detection of pure spin currents, are observed.   
The spin diffusion length and spin lifetime of the Si$_{\rm 0.1}$Ge$_{\rm 0.9}$ layer at low temperatures are reliably estimated to be $\sim$0.5 $\mu$m and $\sim$0.2 ns, respectively. 
This study demonstrates the possibility of exploring physics and developing spintronic applications using SiGe alloys.
\end{abstract}


\maketitle
Binary semiconductor alloys, Si$_{1-x}$Ge$_{x}$ (0 $\le x \le$ 1), have been studied in the field of complementary metal-oxide semiconductor (CMOS) technologies,\cite{Takagi,Mizuno,Paul,Thompson,Lee,Miyao1,Miyao2} optoelectronics for telecommunications,\cite{Kuo,Soref} quantum-Hall systems,\cite{Lai_QH,Hamaya_QH,Ismail} and quantum information processing.\cite{Kawakami1,Takeda}  
In particular, the SiGe alloys have been utilized for introducing the strain to the channel layers in the source/drain area or in the substrate to enhance the electron and hole mobility in CMOS transistors.\cite{Takagi,Mizuno,Paul,Thompson,Lee} 
Also, the Si/SiGe and Ge/SiGe heterostructures enable bandgap engineering and one can control the conduction and valence band structures by adjusting the Ge content $x$.\cite{People,Kris,Fischetti,Paul} Accordingly, if spintronic technologies are integrated into the Si-CMOS technologies, the compatibility between SiGe and spintronics should be explored. 

By electrical means in spin-valve device structures, spin injection/detection and spin relaxation in Si\cite{Appelbaum,Jonker_APL,Ando_APL,Suzuki,Ishikawa,Spiesser,Sasaki} and Ge\cite{Zhou,Kasahara_APEX,Li_PRL,Fujita_PRB,Yamada_PRB,Fujita_PRAP} have been investigated in detail.
In particular, recent progress of technological developments for detecting pure spin current transport at room temperature in Si \cite{Suzuki,Ishikawa,Spiesser} and Ge \cite{Yamada_APEX} by four-terminal nonlocal magnetoresistance measurements is noteworthy.
On the other hand, those technologies in SiGe alloys have not been developed yet although the generation of spin polarized carriers induced by circularly polarized light has been reported.\cite{Lange_PRB,Ferrari_PRB,Giorgioni_APL} 
For bulk Si$_{1-x}$Ge$_{x}$ (0 $\le x \le$ 1), if the composition reaches $x \sim$ 0.85, the bottom of the conduction band can vary from $L$ point to ${\it \Delta}$ one,\cite{People,Kris,Paul,Fischetti} leading to the marked change in electrical properties including $g$-factor.\cite{Vrijen_PRA}  
However, there is almost no information on physics of pure spin current transport in SiGe alloys. 
As a first step for developing SiGe spintronic technologies, one should concentrate on a simple composition of $x >$ 0.85, known to maintain a Ge-like electronic band structure having the bottom of the conduction band at around $L$ point in the ${\bf k}$-space.\cite{People,Kris,Fischetti,Paul}

In this letter, by using four-terminal nonlocal magnetoresistance measurements in Si$_{\rm 0.1}$Ge$_{\rm 0.9}$-based lateral spin-valve (LSV) devices, we show reliable pure spin current transport in an {\it n}-type Si$_{\rm 0.1}$Ge$_{\rm 0.9}$ ({\it n}-SiGe) layer at low temperatures. 
Clear nonlocal spin-valve signals and Hanle-effect curves are observed at low temperatures, indicating generation, manipulation, and detection of pure spin currents in {\it n}-SiGe. From one dimensional spin diffusion models, we can estimate the spin diffusion length ($\lambda_{\rm SiGe}$) and spin lifetime (${\tau_{\rm SiGe}}$) of a SiGe layer used here to be $\sim$0.5 $\mu$m and $\sim$0.2 ns, respectively, at low temperatures. 

In the following, the growth of a Si$_{\rm 0.1}$Ge$_{\rm 0.9}$ spin-transport layer used in this study is explained. 
Using molecular beam epitaxy (MBE), we firstly formed an undoped Ge(111) layer ($\sim$30 nm) grown at 350 $^\circ$C (LT-Ge) on the undoped Si(111) substrate ($\rho$ $\sim$ 1000 $\Omega$cm), followed by an undoped Ge(111) layer ($\sim$70 nm) grown at 700 $^\circ$C (HT-Ge).\cite{Sawano_TSF,Pezzoli_PRAP}   
Next, we grew a 70-nm-thick phosphorous (P)-doped $n$-Si$_{\rm 0.1}$Ge$_{\rm 0.9}$(111) layer by MBE at 350 $^\circ$C on top of the HT-Ge layer. 
The Ge content $x$ in the SiGe layer was adjusted by controlling the deposition rates of Si and Ge, determined by x-ray diffraction (XRD) measurements of the SiGe layer grown on Ge(111) substrate.\cite{Powell} 
The carrier concentration ($n$) in the $n$-SiGe(111) layer was determined to be $n \sim$ 5 $\times$ 10$^{18}$ cm$^{-3}$ from Hall-effect measurements. 
Since the electrical properties of the HT-Ge layer were $p$-type conduction and relatively high resistivity compared to the spin transport ($n$-SiGe) layer, we can ignore the spin diffusion into the HT-Ge layer. 
To promote the tunneling conduction of electron spins through the Schottky barriers,\cite{Hamaya_Gr} a 7-nm-thick P $\delta$-doped Ge layer with an ultra-thin Si layer was grown on top of the spin-transport layer.\cite{MYamada}
A schematic of the grown heterostructure is illustrated in Fig. 1(a). 
\begin{figure}[t]
\begin{center}
\includegraphics[width=8.5cm]{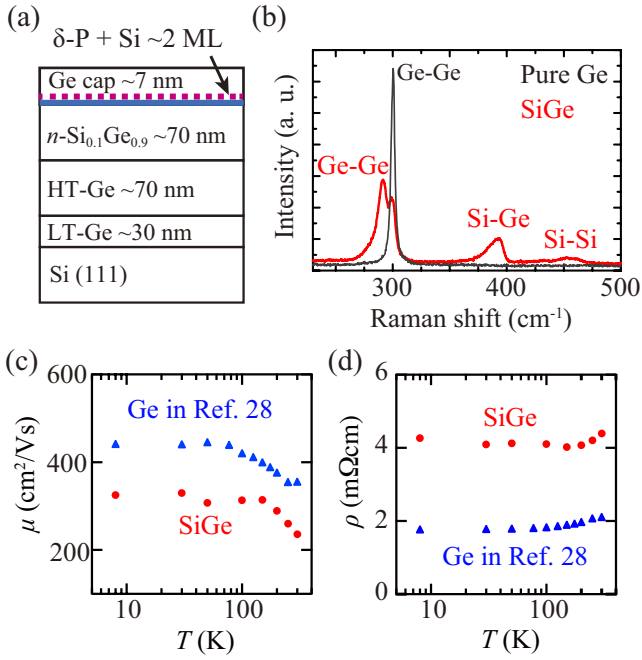}
\caption{(Color online) (a) Schematic of the grown Si$_{\rm 0.1}$Ge$_{\rm 0.9}$/Ge/Si(111) heterostructure for the spin transport measurements in SiGe. (b) Room-temperature Raman spectra of the grown SiGe layer prior to the growth of the Co$_{2}$FeAl$_{0.5}$Si$_{0.5}$ layer and a pure Ge layer. (c) and (d) are $\mu - T$ and  $\rho - T$ plots for the grown SiGe layer, together with the pure Ge layer in Ref. \cite{Fujita_PRB}. }
\end{center}
\end{figure}

Figure 1(b) shows Raman spectra of the grown SiGe layer, together with a pure Ge. 
The three main peaks corresponding to the Si-Si ($\sim$450 cm$^{-1}$), Si-Ge ($\sim$390 cm$^{-1}$), and Ge-Ge ($\sim$290 cm$^{-1}$) bonds are observed, and their positions are consistent with the previous reports for high Ge-content SiGe layers.\cite{Pezzoli, Pezzoli2} 
Here the weaker peak at $\sim$300 cm$^{-1}$, almost equivalent to that from the pure Ge, comes from the 7-nm-thick Ge capping layer in Fig. 1(a). 
From these Raman spectra, we can judge that the (111)-oriented SiGe layer was formed on the epitaxial Ge layer on Si(111).  
In addition, we characterized the grown SiGe layer by XRD and the fringe patterns related to the SiGe peak were not observed (not shown here), indicating that the grown Ge/SiGe interface was not smooth and fully pseudomorphic. Thus, the misfit dislocations were included in the grown SiGe layer. 
From longitudinal and transverse resistance measurements, we confirmed the $n$-type conduction and degenerated electrical properties of the grown SiGe layer in the range of 8 to 300 K, as shown in Figs. 1(c) and 1(d).  
An electron mobility ($\mu$) of $\sim$325 cm$^{2}$/Vs and an $n$ of $\sim$5 $\times$ 10$^{18}$ cm$^{-3}$ were obtained at 8 K, which are not markedly lower than those of pure Ge in our previous works.\cite{Fujita_PRB,Yamada_PRB,Fujita_PRAP} 
Therefore, we can understand that the influence of the misfit dislocations on electrical properties for SiGe layer is relatively limited. 
\begin{figure}[t]
\begin{center}
\includegraphics[width=7cm]{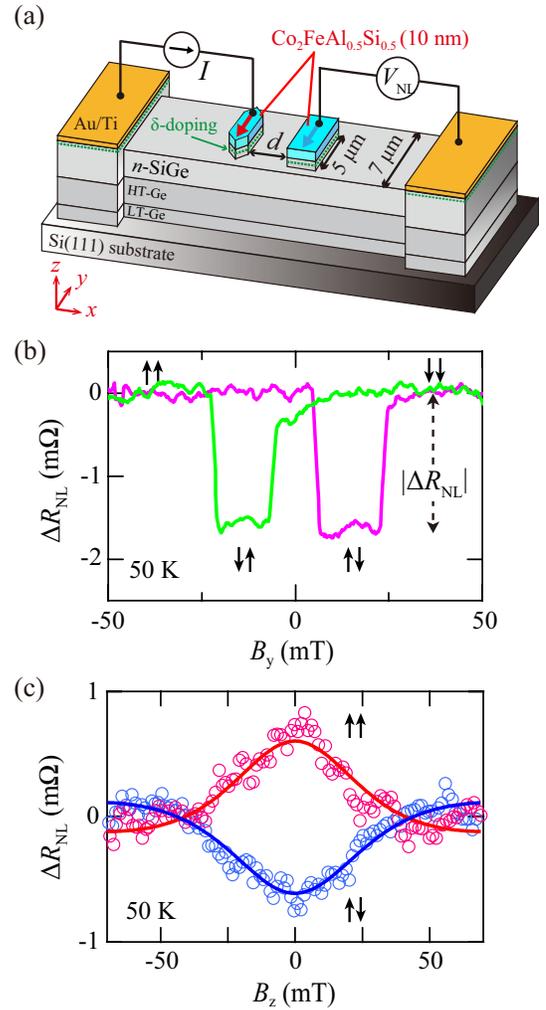}
\caption{(Color online) (a) Schematic illustration of the fabricated CFAS/$n$-SiGe based LSV. (b) NL magnetoresistance curve and (c) Hanle effect curves of the CFAS/$n$-SiGe LSV for the parallel and antiparallel magnetization configurations at 50 K at $I$ $=$ $-$3.5 mA. }
\end{center}
\end{figure}

To investigate spin transport in the grown SiGe layer, we grew a 10-nm-thick Co$_{2}$FeAl$_{0.5}$Si$_{0.5}$ (CFAS) film as a spin injector/detector on top of it by well-established low-temperature MBE techniques.\cite{Fujita_PRAP,SYamada} 
Figure 2(a) shows a schematic of the fabricated lateral spin valves (LSVs) with the CFAS/SiGe Schottky-tunnel contacts. The detailed fabrication processes were described elsewhere.\cite{Fujita_PRAP} 
The sizes of the spin injector and detector in the LSVs are 0.4 $\times$ 5.0 $\mu$m$^{2}$ and 1.0 $\times$ 5.0 $\mu$m$^{2}$, respectively.  
To evaluate the spin diffusion length of the SiGe layer, we changed the edge-to-edge distance ($d$) between the spin injector and spin detector from 0.4 to 2.0 $\mu$m. 
By measuring $I-V$ characteristics, Schottky-tunnel conduction through the CFAS/SiGe interfaces was seen, as shown in our previous works\cite{Yamada_PRB,Fujita_PRAP} and these properties were reproduced from device to device. 

Figure 2(b) displays a hysteresis curve of the nonlocal magnetoresistance ($\Delta R_{\rm NL} =$ $\Delta V_{\rm NL}$/$I$) in an LSV with $d =$ 0.5 $\mu$m by applying in-plane magnetic fields ($B_{\rm y}$), measured at $I =$ -3.5 mA at 50 K, where the negative sign of $I$ ($I <$ 0) means that the electrons are injected from CFAS into the SiGe channel, i.e., spin injection condition, through the Schottky tunnel barrier. 
A nonlocal spin-valve signal with a magnitude ($|\Delta R_{\rm NL}|$) of $\sim$ 1.7 m$\Omega$ can be seen at 50 K, in which the observed spin-valve behavior is attributed to the change in the magnetization configurations of the two different CFAS contacts used between parallel and anti-parallel magnetization states. 
However, the value of $|\Delta R_{\rm NL}|$ is markedly small compared to that for CFAS/Ge-LSVs.\cite{Yamada_PRB,Fujita_PRAP} 
The difference in $|\Delta R_{\rm NL}|$ might be due to the degraded quality of the CFAS/SiGe heterointerface. 
Using the nonlocal four-terminal geometry under applying out-of-plane magnetic field ($B_{\rm  z}$), we also record Hanle-type spin precession curves for the parallel and anti-parallel magnetization states of the CFAS contacts in Fig. 2(c). The recorded $\Delta R_{\rm NL}$ curves are evidence for the generation, manipulation, and detection of pure spin currents in the {\it n}-SiGe layer. These data shown in Fig. 2 mean the first experimental demonstration of the pure spin current transport in a {\it n}-SiGe alloy. 

Using the following one dimensional spin drift diffusion model \cite{Jedema_Nature,Lou_NatPhys}, we can tentatively obtain a spin lifetime ($\tau_{\rm SiGe}$) and a diffusion constant ($D$) of the SiGe layer used here.
\begin{equation}
\Delta R_{\rm NL}(B_{\rm z}) = \pm A{ {\int_0^{\infty}}{\phi(t)}{\rm cos}({\omega}_{L}t){\exp\left(-\frac{t}{\tau_{\rm SiGe}}\right)}dt},
\end{equation}
where $A =$ ${\frac{{P_{\rm inj}}{P_{\rm det}}{\rho_{\rm SiGe}}D}{S}}$ and  $\phi(t) =$ $\frac{1}{\sqrt{4{\pi}Dt}}{\exp\left(-\frac{L^{2}}{4Dt}\right)}$.
$P$$_{\rm inj}$ and $P$$_{\rm det}$ are electron spin polarizations in SiGe created by the spin injector and detector, respectively, $\rho$$_{\rm SiGe}$ is the resistivity ($\rho$$_{\rm SiGe}$ = 4.3 m$\Omega$cm), $S$ is the cross section ($S =$ 0.49 $\mu$m$^{2}$) of the $n$-SiGe layer used here. 
$L$ is the center-to-center distance between the spin injector and detector ($L =$ 1.2 $\mu$m), $\omega$$_{L}$ (= $g$$\mu$$_{\rm B}$$B$$_{z}$/$\hbar$) is the Larmor frequency, $g$ is the electron $g$-factor ($g$ = 1.56) in Si$_{\rm 0.1}$Ge$_{\rm 0.9}$,\cite{Vrijen_PRA}
$\mu$$_{\rm B}$ is the Bohr magneton. 
As a result, a ${\tau_{\rm SiGe}}$ of 0.2 $\pm$ 0.06 ns and a $D$ of 11.2 $\pm$ 0.6 cm$^{2}$/s can be estimated from the fitting to the Hanle data with Eq. (1). 
Using the relation, $\lambda$ $=$ $\sqrt{D\tau_{\rm s}}$, we can roughly obtain a $\lambda_{\rm SiGe}$ of 0.47 $\pm$ 0.02 $\mu$m at 50 K. 

We also measured $\Delta R_{\rm NL}$ for SiGe-LSVs with various $d$. 
The $d$ dependence of $|\Delta R_{\rm NL}|$ at 30 and 50 K is shown in Fig. 3. 
The value of $|\Delta R_{\rm NL}|$ is exponentially decreased with increasing $d$. 
In general, the $d$ dependence can be represented by the following equation \cite{Johnson_PRL, Jedema_Nature},
\begin{equation}
|\Delta R_{\rm NL}| ={ \frac{|{P_{\rm inj}|}|{P_{\rm det}|}{\rho_{\rm SiGe}}{\lambda_{\rm SiGe}}}{S}}{\exp\left(-\frac{d}{\lambda_{\rm SiGe}}\right)},
\end{equation}
where $\rho$$_{\rm SiGe} =$ 4.3 m${\Omega}$cm and $S =$ 0.49 $\mu$m$^{2}$. 
From fitting the decay of $|\Delta R_{\rm NL}|$ to Eq. (2), the values of $\lambda_{\rm SiGe}$ can be estimated to be 0.51 and 0.48 $\mu$m at 30 and 50 K, respectively. 
These values are consistent with the obtained value from the fits to the Hanle data shown in previous paragraph. 
We also note that a $\lambda_{\rm SiGe}$ of $\sim$0.5 $\mu$m at low temperatures is consistent with the spin diffusion length of pure Ge layers with $n =$ 4$-$8 $\times$ 10$^{18}$ cm$^{-3}$ at low temperatures, reported previously in Ref. \cite{Fujita_PRB,Yamada_PRB}.  
Although we could not examine the $d$ dependence of the nonlocal spin signals at higher temperatures, there was almost no change in $\lambda_{\rm SiGe}$ between 30 and 50 K, also consistent with Ge in Ref. \cite{Fujita_PRB}.
\begin{figure}[t]
\begin{center}
\includegraphics[width=8cm]{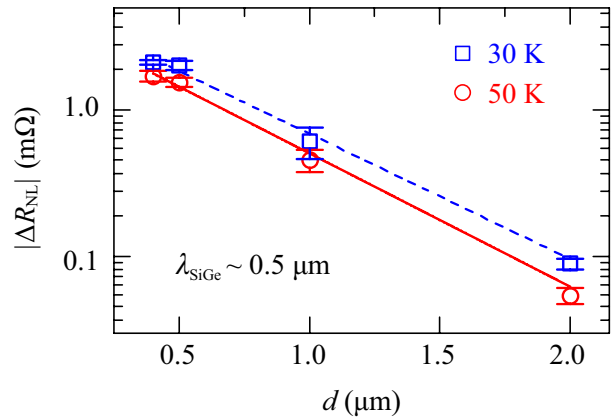}
\caption{(Color online) $d$ dependences of $|\Delta R_{\rm NL}|$ for a SiGe LSV at 30 (blue) and 50 K (red). The solid and dashed lines indicate the results of fitting to Eq. (2). }
\end{center}
\end{figure}
We can also discuss ${\tau_{\rm SiGe}}$ from the $d$ dependence in Fig. 3 and the relation of $\lambda$ $=$ $\sqrt{D\tau_{\rm s}}$. 
Here the value of $D$ can be estimated from Eq. (4) in Ref. \cite{Flatte_PRL} and the electron mobility ($\mu$ $\sim$ 330 cm$^{2}$/Vs at 30 K, $\mu$ $\sim$ 307 cm$^{2}$/Vs at 50 K) of the used SiGe layer, experimentally obtained by Hall-effect measurements.
The calculated values of $D$ at 30 and 50 K are 13.8 cm$^{2}$/s and 13.5 cm$^{2}$/s, respectively. 
As a consequence, ${\tau_{\rm SiGe}} \sim$ 0.18 ns can be obtained at 30 and 50 K.  
The estimated ${\tau_{\rm SiGe}}$ value is also consistent with that obtained from Hanle measurements. 
From these results, we can judge that, for $n$-Si$_{\rm 0.1}$Ge$_{\rm 0.9}$, the reliable values of $\lambda_{\rm SiGe}$ ($\sim$0.5 $\mu$m) and ${\tau_{\rm SiGe}}$ ($\sim$ 0.2 ns) at low temperatures are shown in this study. 

Recently, Song {\it et al.} theoretically proposed that the spin relaxation at low temperatures in multivalley semiconductors such as Si and Ge is dominated by the intervalley spin-flip scattering induced by the central-cell potential of impurities,\cite{Song_PRL} which is so called the donor-driven spin relaxation. 
We have experimentally clarified that the spin relaxation mechanism at low temperatures in degenerate Si\cite{Ishikawa} and Ge\cite{Fujita_PRB,Yamada_PRB} cannot quantitatively be interpreted in terms of the Elliott-Yafet mechanism but the mechanism due to the impurity- and phonon-induced intervalley spin-flip scattering.  
Using their theory,\cite{Song_PRL} we can tentatively discuss the spin lifetime in the degenerate SiGe layers. 
When the Fermi energy ($\epsilon_{\rm F}$) is larger than the thermal energy ($k_{\rm B}T$) at low temperatures, the spin scattering rate ($\frac{1}{\tau}$) depends on the concentration of the donor impurity ($N_{\rm d}$) in degenerate conditions,\cite{Song_PRL} where $k_{\rm B}$ is the Boltzmann's constant. 
If $N_{\rm d}$ is regarded as $n$ in the used degenerate Si$_{\rm 0.1}$Ge$_{\rm 0.9}$, the donor-driven spin relaxation in degenerate SiGe can be expressed as follows. 
\begin{equation}
\frac{1}{\tau_{\rm }} \approx \frac{4{\pi}nm_{\rm e}a_{\rm B}^6}{27\hbar^{3}}\left(3\pi^{2}n\right)^{\frac{1}{3}}\Delta_{\rm so}^{2},
\end{equation}
where $a_{\rm B}$ and $m_{\rm e}$ are the Bohr radius and the electron effective mass in Si$_{\rm 0.1}$Ge$_{\rm 0.9}$, respectively. $\Delta_{\rm so}$ is the spin-orbit coupling induced splitting of the triply degenerated 1$s$ ($T$$_{\rm 2}$) donor state in Si$_{\rm 0.1}$Ge$_{\rm 0.9}$. 
Here we assigned $\epsilon_{\rm F} \approx \frac{\hbar^{2}}{2m_{e}}\left(3\pi^{2}n\right)^{2/3}$ to conduction electron energy ($\epsilon_{\bf k}$) from Eq. (4) in Ref. \cite{Song_PRL}. 
According to previous literature,\cite{Vrijen_PRA,People,Kris,Paul} we can assume that the parameters for Si$_{\rm 0.1}$Ge$_{\rm 0.9}$ such as $a_{\rm B}$, $m_{\rm e}$, $\Delta_{\rm so}$ are almost equivalent to those in pure Ge. Thus, we tentatively used $a$$_{\rm B}$ $=$ 6.45 nm,\cite{Wilson_PR} $n =$ 5.0 $\times$ 10$^{18}$ cm$^{-3}$, $m$$_{\rm e}$ $=$ 0.16$m$$_{\rm 0}$,\cite{Spitzer_JAP} and $\Delta$$_{\rm so}$ $=$ 0.11 meV.\cite{Fujita_PRB,Yamada_PRB} Here the assumed $\Delta$$_{\rm so}$ values are much smaller than the valley-orbit induced singlet-triplet splitting in P doped Ge of $\sim$2.83 meV \cite{Reuszer_PR}. 
As a result, we can obtain a spin lifetime of $\sim$0.3 ns, nearly consistent with the experimentally estimated values shown in the previous paragraph. 
These mean that the spin-related physics including spin relaxation mechanism in Si$_{\rm 0.1}$Ge$_{\rm 0.9}$ is similar to that in pure Ge.\cite{Fujita_PRB,Yamada_PRB,Fujita_PRAP}
From now on, if Si$_{1-x}$Ge$_{x}$ alloys with $x \le$ 0.85 is utilized as a spin transport layer, pure spin current transport showing characteristics of Si-like electronic band structures may be observed. 
In our experiments, however, the growth of the Si$_{1-x}$Ge$_{x}$ alloys is limited only on the HT-Ge(111)/LT-Ge(111) structure on Si(111) substrate, leading to the strained Si$_{1-x}$Ge$_{x}$ alloys with $x \le$ 0.85. 
When the strain is induced in the Si$_{1-x}$Ge$_{x}$ alloys, the strain effect on the spin-related physics can be dominant.\cite{Chalaev_PRB}  
Therefore, at around $x \sim$ 0.85, it may be difficult to observe the great change in the spin-related physics depending on $x$. 

In summary, we studied pure spin current transport in Si$_{\rm 0.1}$Ge$_{\rm 0.9}$ alloy ($n \sim$ 5.0 $\times$ 10$^{18}$ cm$^{-3}$). 
Four-terminal nonlocal magnetoresistance signals and Hanle-effect curves  were observed in the SiGe-based LSVs. 
The spin diffusion length and spin lifetime of a SiGe layer at low temperatures were experimentally estimated to be $\sim$0.5 $\mu$m and $\sim$0.2 ns, respectively. 
This study demonstrates the possibility of exploring physics and developing spintronic applications using SiGe alloys.

This work was partly supported by JSPS KAKENHI (No. 16H02333, 17H06120, 17H06832) and a Grant-in-Aid for Scientific Research on Innovative Areas `Nano Spin Conversion Science' (No. 26103003) from MEXT. 
M.Y. acknowledges Scholarships from Toyota Physical and Chemical Research Institute Foundation.  


\end{document}